\documentclass[aps,pra,groupedaddress,twocolumn]{revtex4}
\usepackage{graphicx}
\usepackage{amsmath}
\usepackage{amsxtra}
\usepackage{amssymb}
\input epsf

\newcommand{\bra}[1]{\langle{#1}|}
\newcommand{\ket}[1]{|{#1}\rangle}

\newcommand{\projektor}[1]{\ket{#1}\bra{#1}}

\newcommand{\ketbra}[2]{|{#1}\rangle \langle{#2}|}

\newcommand{\Trace}{\operatorname{Tr}}

\begin{document}

\title{Phase-dependent light propagation in atomic vapors}

\author{Sarah Kajari-Schr\"oder,$^1$ Giovanna Morigi,$^2$
Sonja Franke-Arnold,$^{3,4}$ and Gian-Luca Oppo$^4$}
\affiliation{$^1$ Institut f\"ur Quantenphysik, University of Ulm, D-89069
Ulm, Germany\\
$^2$ Grup d'Optica, Departament de Fisica, Universitat Autonoma de
Barcelona, 08193 Bellaterra, Spain\\
$^3$ Department of Physics, University of Glasgow, G12 8QQ
Glasgow, Scotland, U.K.\\
$^4$ Department of Physics, University of Strathclyde, G14 0NG Glasgow,
Scotland, U.K.}

\date{\today}

\begin{abstract} Light propagation in an atomic medium whose
coupled energy levels form a $\diamondsuit$-configuration exhibits
a critical dependence on the input conditions. Depending on the
relative phase of the input light fields, the response of the
medium can be dramatically modified and switch from opaque to
semi-transparent. These different types of behaviour are caused by
the formation of coherences due to interference in the atomic
excitations. Alkali-earth atoms with zero nuclear spin are ideal
candidates for observing these phenomena which could offer new
perspectives in control techniques in quantum electronics.
\end{abstract}

\maketitle

\section{Introduction}

Experimental evidence has demonstrated that the nonlinear optical
properties of laser-driven atomic gases exhibit counter-intuitive
features with promising applications. A peculiarity of these media
is the possibility to manipulate their internal and external
degrees of freedom with a high degree of control. Recently the
control of the internal dynamics in an atomic vapor by means of
electromagnetically induced transparency (EIT)~\cite{EIT} was
demonstrated for the generation of four-wave mixing
dynamics~\cite{Harris04} and of controlled quantum pulses of
light~\cite{Harris05,Lukin05}. Zeeman coherence has also been used
to induce phase dependent amplification without inversion in
Samarium vapors~\cite{Nottelmann93} and in HeNe
mixtures~\cite{Peters96}. In another experiment, the interplay of
internal and external degrees of freedom in an ultracold atomic
gas by means of recoil-induced resonances~\cite{RIR} was used to
achieve waveguiding of light~\cite{Prentiss05}. From this
perspective, it is important to identify further possible control
parameters on the atomic dynamics for the manipulation of the
non-linear optical response of the medium.

Recent studies have been focusing on the dynamics of light
interacting with atoms featuring coupled energy levels in a
so-called 'closed-loop' configuration~\cite{Buckle86,Kosachiov92}.
In this configuration a set of atomic states is (quasi-)
resonantly coupled by laser fields so that each state is connected
to any other via two different paths of coherent
photon-scattering. As a consequence, the relative phase between
the transitions critically influences dynamics~\cite{Buckle86} and
steady states~\cite{Kosachiov92,Korsunsky99,Morigi02}.
Applications of closed-loop configurations to nonlinear optics
have featured double-$\Lambda$ systems where two stable or
metastable states are -each- coupled to two common excited states.
A rich variety of nonlinear optical phenomena has been
predicted~\cite{Korsunsky99,PhaseoniumRev,Wilson-Gordon} and
experimentally observed~\cite{VandenHeuwell,Nottelmann93,Peters96,
Windholz96,Harris00,Harris04,Windholz99,Windholz04}.
In~\cite{Windholz04}, in particular, it has been shown
experimentally that the properties of closed-loop configurations
can be used to correlate electromagnetic fields with carrier
frequency differences beyond the GHz regime. Moreover, coherent
control based on the relative phase in closed-loop configuration
has been proposed in the context of quantum information
processing~\cite{Sola05}.

\begin{figure}[h] \begin{center}
\includegraphics[width=5cm]{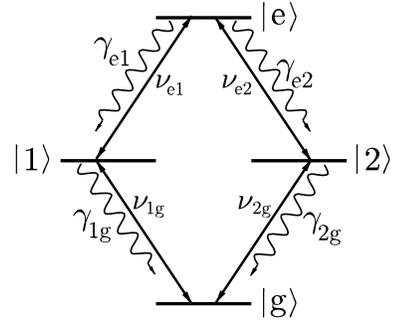} \caption{Electronic
transitions of the $\diamondsuit$-configuration. Each transition
$|i\rangle\to|j\rangle$ is resonantly driven by a laser field at
frequency $\nu_{ij}$. Here, $|g\rangle$ is the ground state,
$|1\rangle$ and $|2\rangle$ the intermediate states, which decay
into the ground state at rates $\gamma_{1g}$ and $\gamma_{2g}$,
respectively, and $|e\rangle$ the excited state, which decays with
rates $\gamma_{e1}$ and $\gamma_{e2}$ into the corresponding
intermediate states. Each pair of levels is coupled by two paths
of excitation, hence the dynamics depends critically on the
relative phase between the paths. The coherent dynamics of the
$\diamondsuit$-configuration is equivalent to that of the
double-$\Lambda$ scheme, whereas the radiative instability of the atomic 
levels differs. } \label{Fig:1} \end{center} \end{figure}

In this work we investigate the phase-dependent dynamics of light
propagation in a medium of atoms whose energy levels are driven in
a closed-loop configuration, denoted by the $\diamondsuit$
(diamond) scheme and depicted in Fig.~\ref{Fig:1}. This
configuration consists of four driven transitions where one ground
state is coupled in a V-type structure to two intermediate states,
which are in turn coupled to a common excited state in a
$\Lambda$-type structure. It can be encountered, for instance, in
(suitably driven) isotopes of alkali-earth atoms with zero nuclear
spin~\cite{AlkaliEarth}. Although the coherent dynamics of
$\diamondsuit$ schemes is equivalent to that of double-$\Lambda$
systems~\cite{Buckle86}, the steady states of the two systems
exhibit important differences due to the different relaxation
processes~\cite{Korsunsky99,Morigi02}.

The dynamics of light propagation in a medium of
$\diamondsuit$-atoms is studied by integrating numerically the
Maxwell-Bloch equations. We find that, depending on the input
field parameters, the polarization along the medium can be
drastically modified. The propagation dynamics may exhibit two
metastable values of the relative phase, namely the values $0$ and
$\pi$, corresponding to a semi-transparent and to an opaque
medium, respectively. For different values of the initial phase,
light propagation along the medium tends to one of these two
values, depending on the input values of the driving amplitudes.
These two types of the medium response are supported by the
formation of atomic coherences leading to a minimization of
dissipation by depleting the population of one or more atomic
states. This phase dependent behavior, selected at the input by
the operator, offers promising perspectives in control techniques
in quantum electronics.

The article is organized as follows. In Sec.~\ref{Sec:Model} the
model is introduced and discussed. In Sec.~\ref{Sec:Propagation}
the results for the dynamics of light propagation, solved
numerically from the equations reported in
Sec.~\ref{Sec:Equations}, are reported and discussed in some
parameter regimes. Conclusions and outlooks are reported in
Sec.~\ref{Sec:Conclude}. The appendices present in detail
equations and calculations at the basis of the model derived in
Sec.~\ref{Sec:Model}.

\section{The Model}
\label{Sec:Model}

We consider a classical field propagating in a dilute atomic gas
along the positive $z$-direction. The field is composed of four
optical frequencies $\nu_{1g}$, $\nu_{2g}$, $\nu_{e1}$ and
$\nu_{e2}$, its complex amplitude is a function of time $t$ and
position $z$ of the form \begin{equation} {\bf
E}(z,t)=\frac{1}{2}\sum_{i,j} \mathcal{E}_{ij}(z,t){\bf e}_{ij}
\text{e}^{-i(\nu_{ij} t-k_{ij}z+\phi_{ij}(z,t))} +{\rm c.c.},
\label{Fields} \end{equation} where $k_{ij}$ denotes the wave
vector and ${\bf e}_{ij}$ the polarization of the frequency
component $\nu_{ij}$. The input field enters the medium at $z=0$,
and the effect of coupling to the medium is accounted for in the
$z$ dependence of the amplitude $\mathcal{E}_{ij}(z,t)$ and phase
$\phi_{ij}(z,t)$ whose variations in position and time are slow
with respect to the wavelengths $\lambda_{ij}=2\pi/k_{ij}$ and the
oscillation periods $T=2\pi/\nu_{ij}$, respectively. The atomic
gas is very dilute and we can assume that the atoms interact with
the fields individually. In particular, each field component at
frequency $\nu_{ij}$ drives (quasi-) resonantly the electronic
transition $\ket{i}\to\ket{j}$ of the atoms in the medium, such
that the atomic levels are coupled in a $\diamondsuit$-shaped
configuration.

The relevant atomic transitions and the coupling due to the lasers
are displayed in Fig.~\ref{Fig:1}.  The ground state $\ket{g}$ is
coupled to the intermediate states $\ket{1}$, $\ket{2}$ at
energies $\hbar\omega_1, \hbar \omega_2$ by transitions with
dipole moments ${\bf d}_{1g}=\bra{1}{\bf d}\ket{g}$ and ${\bf
d}_{2g}=\bra{2}{\bf d}\ket{g}$, respectively. The intermediate
states decay back into the ground state at decay rates
$\gamma_{1g}$ and $\gamma_{2g}$. The intermediate states are also
coupled to the excited state $\ket{e}$ at an energy $\hbar
\omega_e$ with respect to the ground state $\ket{g}$, by the
dipole transitions ${\bf d}_{e1}=\bra{e}{\bf d}\ket{1}$, ${\bf
d}_{e2}=\bra{e}{\bf d}\ket{2}$. The excited state $|e\rangle$
decays into states $|1\rangle$ and $|2\rangle$ at rates
$\gamma_{e1}$ and $\gamma_{e2}$, respectively. A similar
configuration of levels can be found in isotopes of alkali atoms
with zero nuclear spin~\cite{AlkaliEarth}.

The light fields propagating through the dilute atomic sample will
induce a macroscopic polarization in the atoms.  This polarization
will depend on intensities and phases of the light fields.  The
polarization, in turn, will affect absorption and refraction of the
light fields, altering their propagation. Below we introduce the
equations for field propagation and the corresponding atomic
dynamics.

\subsection{Equations for field propagation}

We denote by ${\bf P}(z,t)$ the macroscopic polarization induced
in the atomic gas \begin{eqnarray} {\bf P}(z,t) =n\Trace\{{\bf
\hat{d}}\sigma(z,t)\} \label{polarisation} \end{eqnarray} where
${\bf \hat{d}}$ is the dipole operator, $n$ is the density of the
medium, which we assume to be zero for $z<0$ and uniform for
$z>0$, and $\sigma(z,t)$ is the atomic density matrix at time $t$
and position $z$, which has been obtained by tracing out the other
external degrees of freedom. Details of the underlying assumptions
at the basis of Eq.~(\ref{polarisation}) are discussed in
Appendix~A.

We decompose the polarization ${\bf P}(z,t)$ into slowly- and
fast-varying components, namely \begin{equation} {\bf
P}(z,t)=\frac{1}{2}\sum_{i,j} \mathcal{P}_{ij}(z,t) {\bf e}_{ij}
\text{e}^{ -i(\nu_{ij} t-k_{ij}z+\phi_{ij}(z,t))} +{\rm c.c.},
\label{polallgemein} \end{equation} whereby the complex amplitudes
$\mathcal{P}_{ij}$ and the phases $\phi_{ij}$ vary slowly as a
function of position and time. We consider the parameter regime
where the driving fields are sufficiently weak so that the
generation of higher-order harmonics can be neglected. By
comparing Eqs.~(\ref{polarisation}) and~(\ref{polallgemein}), the
amplitudes $\mathcal{P}_{ij}$ can be expressed in terms of the
elements of the atomic density matrix $\sigma$, \begin{equation}
\mathcal{P}_{ij}=2n \mathcal{D}_{ij} \sigma_{ij}e^{i(\nu_{ij}
t-k_{ij}z+\chi_{ij})} \end{equation} where
$\sigma_{ij}=\bra{i}\sigma\ket{j}$. We have expressed the dipole
moments in direction of the electric field polarization as ${\bf
e}_{ij}\cdot {\bf d}_{ji}=\mathcal{D}_{ij}{\rm e}^{-{\rm
i}\theta_{ij}}$, thereby separating the complex amplitudes
$\mathcal{P}_{ij}$ into modulus and phase. Here, the term
$\mathcal{D}_{ij}$ is real, $\theta_{ij}$ are the dipole phases
($\theta_{ij}=-\theta_{ji}$), and \begin{equation} \label{chi:ij}
\chi_{ij}(z,t)=\phi_{ij}(z,t)-\theta_{ij} \end{equation} is the
sum of the slowly-varying field phases $\phi_{ij}(z,t)$ and the
dipole phases $\theta_{ij}$.

Using definitions~(\ref{Fields}) and~(\ref{polallgemein}) and applying
a coarse-grained description in time and space, the Maxwell equations
simplify to a set of propagation equations for each of the slowly-varying
components of the laser and polarization fields~\cite{QScully}
\begin{align} \frac{\partial
\mathcal{E}_{ij}}{\partial z}
  +\frac{1}{c}\frac{\partial \mathcal{E}_{ij}}{\partial t}
  &=-\frac{\nu_{ij}}{2\:\epsilon_0 \:c} {\rm Im} \{\mathcal{P}_{ij}
  (\mathcal{E}_{kl},\phi_{kl})\},\label{feldprop}\\
\frac{\partial \phi_{ij}}{\partial z}
  +\frac{1}{c}\frac{\partial \phi_{ij}}{\partial t}
  &=-\frac{\nu_{ij}}{2\:\epsilon_0\:c}\frac{1}{\mathcal{E}_{ij}}
  {\rm Re} \{\mathcal{P}_{ij}(\mathcal{E}_{kl},\phi_{kl})\},
  \label{phiprop}
\end{align} which are defined for $z>0$. Here, each amplitude
$\mathcal{E}_{ij}$ and phase $\phi_{ij}$ is coupled via the
corresponding polarization $\mathcal{P}_{ij}$ to all other field
amplitudes and phases.

We rescale the propagation equations using the dimensionless
length and time \begin{eqnarray} \xi =\kappa_{1g} z,~~~
\tau=c\:\kappa_{1g} t. \end{eqnarray} Here $\kappa_{1g}$ is the
absorption coefficient \begin{equation} \kappa_{1g}=n
\frac{1}{\gamma_{1g}} \frac{\nu_{1g}\mathcal{D}_{1g}^2 }
{c\epsilon_0 \hbar }\nonumber \end{equation} such that
$1/\kappa_{1g}$ determines the characteristic length at which
light driving the transition $\ket{g}\to\ket{2}$ penetrates a
medium with density $n$. We denote the dimensionless field
amplitudes by \begin{equation} \label{Gij}
\mathcal{G}_{ij}=\frac{\Omega_{ij}}{ \gamma_{1g}}
 \frac{\mathcal{D}_{1g}^2 \nu_{1g}}{\mathcal{D}_{ij}^2 \nu_{ij}},
\end{equation} where \begin{equation} \label{Rabi}
\Omega_{ij}(z,t)=\mathcal{D}_{ij}\mathcal{E}_{ij}(z,t)/\hbar
\end{equation} is the real valued Rabi frequency for the
transition $\ket{i}\rightarrow \ket{j}$. In this notation the
propagation Eqs.~(\ref{feldprop}) and (\ref{phiprop}) reduce to
the form \begin{align} \frac{\partial \mathcal{G}_{ij}}{\partial
\xi}
  +\frac{\partial \mathcal{G}_{ij}}{\partial \tau}&=-
  \; {\rm Im}\{p_{ij}\},\label{edimlos}\\
\frac{\partial \phi_{ij}}{\partial \xi}+\frac{\partial \phi_{ij}}
{\partial \tau}&=- \; \frac{1}{\mathcal{G}_{ij}} \;
  {\rm Re} \{p_{ij}\},\label{phidimlos}
\end{align} where \begin{equation} \label{tilde:sigma}
p_{ij}(\xi,\tau)=\sigma_{ij}\text{exp}\left[i\left(\frac{\nu_{ij}}
{c\;\kappa_{1g}}\tau-\frac{k_{ij}}{\kappa_{1g}}\xi+\chi_{ij})\right)\right]
\; \end{equation} denotes the atomic density matrix elements in a
rotated reference frame. In the remainder of this paper we
consider laser field geometries where $\ket{1}$ and $\ket{2}$ are
states of the same hyperfine multiplet so that
$\nu_{1g}\simeq\nu_{2g}$ and $\nu_{e1}\simeq\nu_{e2}$.

\subsection{Atomic dynamics}

The time evolution of the density matrix $\sigma(z,t)$ for the
atomic internal degrees of freedom at position $z>0$ is governed
by the master equation
\begin{equation} \label{Master:z}
\dot{\sigma}=\frac{1}{{\rm i} \hbar}\left[H(z,t),\sigma
\right]+\mathcal{L}\sigma. \end{equation} where $z$ is a classical
variable. Equation~(\ref{Master:z}) is obtained by tracing out the
degrees of freedom of momentum and of position in the transverse
plane, in the limit in which the medium is homogeneously broadened
and the atoms are sufficiently hot and dilute such that their
external degrees of freedom can be treated classically. Details of
the assumptions at the basis of Eq.~(\ref{Master:z}) are reported
in Appendix A. Here the Hamiltonian \begin{eqnarray} H(z,t)
&=&\sum_{j=e,1,2,g} \hbar \omega_j \projektor{j}\label{hamilton1}\\
&-&\frac{\hbar}{2}\sum_{j=1,2}\Bigl(\Omega_{jg}(z,t) \;
e^{-i(\nu_{jg}t-k_{jg}z+\chi_{jg}(z,t))}\;\ketbra{j}{g}\nonumber\\
& &+\Omega_{ej}(z,t) \; e^{-i(\nu_{ej}t-k_{ej}z+\chi_{ej}(z,t))}\;
\ketbra{e}{j}+{\rm H.c.}\Bigr)\nonumber \end{eqnarray} describes
the coherent dynamics of the internal degrees of freedom, and it
depends on $z$ through the (real-valued) Rabi frequency
$\Omega_{ij}(z,t)$ given in Eq.~(\ref{Rabi}), and through the
field and dipole phases, Eq.~(\ref{chi:ij}).

The states $\ket{1},\ket{2}$ and $\ket{e}$ are unstable and decay
radiatively with rates $\gamma_{1g}$, $\gamma_{2g}$ and
$\gamma_e=\gamma_{e1}+\gamma_{e2}$, respectively. The relaxation
processes are described by \begin{eqnarray} \label{Liou}
\mathcal{L}\sigma&=&\!\!\sum_{j=1,2} \frac{\gamma_{jg}}{2}
\left(2\ketbra{g}{j}\sigma\ketbra{j}{g}-
\ketbra{j}{j}\sigma-\sigma\ketbra{j}{j}\right)\nonumber\\
&+&\!\!\!\! \sum_{j=1,2}\!\! \frac{\gamma_{ej}}{2}
\left(2\ketbra{j}{e}\sigma\ketbra{e}{j}-
\ketbra{e}{e}\sigma-\sigma\ketbra{e}{e}\right), \end{eqnarray}
where the recoil due to spontaneous emission is neglected since
the motion is treated classically. In the remainder of this paper
we assume a symmetrical decay of the excited level,
$\gamma_{e1}=\gamma_{e2}=\gamma_e/2$.

We note that the transitions $|g\rangle\to |j\rangle$ ($j=1,2$)
are saturated when $\Omega_{jg}\ge\gamma_{jg}.$ Correspondingly,
the upper transitions $|j\rangle\to |e\rangle$ are saturated when
$\Omega_{ej}\ge\gamma_{e}+\gamma_{jg}.$ For later convenience, we
introduce \begin{equation} \label{G_e}
\mathcal{\tilde{G}}_{ej}=\frac{\mathcal{G}_{ej}}{1+\gamma_e /
\gamma_{jg}}, \end{equation} which explicitly shows the scalings
of the upper field amplitudes with the corresponding decay rates.

\subsection{The relative phase}

In so-called {\it closed-loop} configurations, like the
$\diamondsuit$ scheme, transitions between each pair of electronic
levels are characterized by -at least- two excitation paths,
involving different intermediate atomic
levels~\cite{Kosachiov92,PhaseoniumRev}. In the $\diamondsuit$
scheme the relative phase between these excitation paths
critically determines the solution of the master equation, and
hence the atomic response during propagation. The role of the
relative phase in the atomic response is better unveiled by moving
to a suitable reference frame for the atomic evolution, which is
defined when all amplitudes $\mathcal{E}_{ij}$ are nonzero.

We denote by $\rho$ the density matrix in this reference frame,
obeying the master equation \begin{equation} \label{master:phase}
\dot{\rho}=\frac{1}{i\hbar} \left[ \tilde{H},\rho \right] +
\mathcal{L}\rho. \end{equation} In this reference frame the
Hamiltonian (\ref{hamilton1}) is transformed
to~\cite{Buckle86,Morigi02} \begin{eqnarray} \label{Ham:phase}
\tilde{H} &=&\hbar \Delta_e \projektor{e}+\hbar \Delta_1
\projektor{1}
+\hbar \Delta_2 \projektor{2}\\
&-&\frac{\hbar}{2}\left( \Omega_{e1}\, \ketbra{e}{1} +\Omega_{e2}\, {\rm e}^{{\rm i}
\Theta(z,t)}\,\ketbra{e}{2} \right.
\nonumber\\
& &\left.+\Omega_{1g}\, \ketbra{1}{g} +\Omega_{2g}\, \ketbra{2}{g}
+{\rm H.c.} \right), \nonumber \end{eqnarray} with the detunings
\begin{align}
\Delta_1&=\omega_1-\nu_{1g},\\
\Delta_2&=\omega_2-\nu_{2g},\\
\Delta_e&=\omega_e-\nu_{e1}-\nu_{1g}. \end{align} The
Hamiltonian~(\ref{Ham:phase}) exhibits an explicit dependence on
the phase \begin{equation} \Theta(z,t)=\Delta \nu \;t-\Delta k \;z
+\Delta \chi(z,t). \end{equation} where \begin{align}
\Delta \nu&=\nu_{e1}+\nu_{1g}-\nu_{2g}-\nu_{e2}, \label{vierphotdet} \\
\Delta k&=k_{e1}+k_{1g}-k_{2g}-k_{e2},  \label{wavevecmis} \\
\Delta\chi(z,t) &=
\chi_{e1}(z,t)+\chi_{1g}(z,t)-\chi_{e2}(z,t)-\chi_{2g}(z,t),
\end{align} with $\chi_{ij}$ as defined in Eq.~(\ref{chi:ij}).

The four-photon detuning $\Delta \nu$ results in a time-dependent
phase, the wave-vector mismatch $\Delta k$ in a position dependent
phase, and $\Delta \chi(z,t)$ comprises the relative dipole and
field phases. In~\cite{Korsunsky99,Morigi02} it has been discussed
how $\Theta(z,t)$ affects the dynamics and steady state of the
atom. The latter exists for $\Delta\nu=0$ and in the remainder of
this article we assume $$\Delta\nu=0,~\Delta k=0,$$ i.e., the
atoms are driven at four-photon resonance and by copropagating
laser fields, such that the wave vector mismatch is negligible.
Hence, the phase $$\Theta(z,t)=\Delta\chi(z,t) $$ depends solely
on the relative dipole phase, which is constant, and on the
relative phase of the propagating fields, which evolves according
to the coupled Eqs.~(\ref{edimlos}) and~(\ref{phidimlos}).

\subsection{Propagation of the field amplitudes and phases}
\label{Sec:Equations}

Having introduced the basic assumptions and definitions, we now
report the equations for the propagation of the field amplitudes
and phases in the $\diamondsuit$--medium, which are numerically
solved in Sec.~\ref{Sec:Propagation}.  We relate the elements of
the density matrix $\rho$ in the new reference frame with the
elements $p_{ij}$ from eq.~(\ref{tilde:sigma}) by
$\rho_{g1}=p_{g1}$, $\rho_{g2}=p_{g2}$, $\rho_{e1}=p_{e1}$, and
\begin{eqnarray*} \rho_{e2}=p_{e2}\text{exp}\left(-{\rm
i}\Theta\right). \end{eqnarray*} The propagation equations for the
light fields in the new reference frame can then be obtained from
Eqs.~(\ref{edimlos})-(\ref{phidimlos}) and take the form
\begin{eqnarray} &&\frac{\partial \mathcal{G}_{jg}}{\partial \xi'}
  =-{\rm Im}\{\rho_{jg}\}\, , \label{epropa:g}\\
&&\frac{\partial \phi_{jg}}{\partial \xi'}=- \frac{{\rm Re}\{\rho_{jg}\}}
{\mathcal{G}_{jg}}
\label{phipropa:g}
\end{eqnarray}
for $j=1,2$ and
\begin{eqnarray}
&&\frac{\partial \mathcal{G}_{e1}}{\partial \xi'}
  =-{\rm Im}\{\rho_{e1}\}\, , \label{epropa:e1}\\
&&\frac{\partial \phi_{e1}}{\partial \xi'}=-
  \frac{{\rm Re}\{ \rho_{e1}\}}{\mathcal{G}_{e1}}\, ,
  \label{phipropa:e1}\\
&&\frac{\partial \mathcal{G}_{e2}}{\partial \xi'}
  =- {\rm Im}\{\rho_{e2}e^{i\Theta}\}\, , \\
&&\frac{\partial \phi_{e2}}{\partial \xi'}=-
  \frac{{\rm Re} \{\rho_{e2}e^{i\Theta}\}}{\mathcal{G}_{e2}}\, ,
\label{phipropa:e2} \end{eqnarray} where we have introduced the
variable $$\xi'=\xi+\tau.$$ These equations describe the evolution
of field amplitudes and phases as a function of the atomic density
matrix elements $\rho_{ij}$. In turn, the values of $\rho_{ij}$
depend on the field amplitudes and the relative phase $\Theta$
according to Eqs.~(\ref{master:phase}) and~(\ref{Ham:phase}). The
propagation dynamics now can be investigated by solving the
coupled Eqs.~(\ref{master:phase}) and
(\ref{epropa:g})-(\ref{phipropa:e2}). The optical Bloch equations
for the density matrix $\rho$ are presented in Appendix~B.

In general, the density matrix elements entering
Eqs.~(\ref{epropa:g})-(\ref{phipropa:e2}) are time dependent, i.e.,
$\rho=\rho(\tau)$. In this paper we consider the case of sufficiently
long laser pulses, such that the characteristic time of change of
amplitude and phase of the fields and the interaction time between light
and atoms exceed the time scale in which the atom reaches the internal
steady state. In this regime, we can neglect transient effects, and the
density matrix elements entering Eq.~(\ref{epropa:g})-(\ref{phipropa:e2})
are the stationary solutions of Eq.~(\ref{master:phase}) satisfying
$\partial\rho/\partial t=0.$ This assumption allows us to neglect the
time derivative in Eqs.~(\ref{epropa:g})-(\ref{phipropa:e2}), hence
taking $\xi'\approx\xi$.

The numerical study of the solutions of
Eqs.~(\ref{epropa:g})-(\ref{phipropa:e2}) presented in this paper
is restricted to certain parameter regimes that single out the
role played by the phase and the radiative decay processes in the
dynamics. In particular, we consider the situation where each
atomic transition is driven at resonance, namely $$\Delta_i=0,$$
for $i=1,2,e$.  Moreover, we restrict ourselves to the regime
where the fields are initially driving the corresponding
transitions at saturation. This latter assumption is important to
guarantee a finite occupation of the excited state $|e\rangle$,
and thus to highlight the dependence of the dynamics on the
relative phase $\Theta$.

During propagation, it may occur that one of the field amplitudes
vanishes in just one point of the propagation variable $\xi'$.
When this happens, the relative phase $\Theta$ is not defined and
its value has to be reset manually by imposing continuity of the
trajectory, when integrating the field equations in amplitude and
phase (see Eqs.~(\ref{epropa:g})-(\ref{phipropa:e2})). The
correctness of this procedure has been checked by comparing the
results with those obtained by integrating the field equations for
the real and imaginary parts of the complex field amplitudes.

\section{Light propagation in the $\Diamond$-medium}
\label{Sec:Propagation}

In this section we summarize some peculiar properties of the
$\diamondsuit$-level scheme, which have been extensively discussed
in~\cite{Morigi02}. These properties provide an important insight
into the propagation dynamics, which we study by solving
numerically the Maxwell-Bloch Equations, Eqs.~(\ref{master:phase})
and~(\ref{epropa:g})-(\ref{phipropa:e2}), in the regime where the
input fields couple resonantly and saturate the corresponding
electronic transitions, as described in Sec.~\ref{Sec:Equations}.

\subsection{Symmetries of the $\diamondsuit$--level scheme}
\label{Sec:Symmetries}

\begin{figure*}[t] \begin{center}
\includegraphics[height=15cm,angle=270]{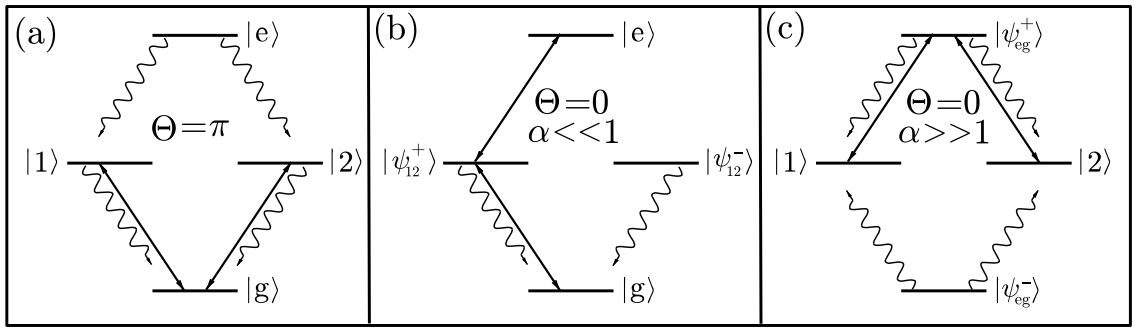}
\caption{Level schemes onto which the $\diamondsuit$ configuration
is mapped, if the lower field amplitudes, as well as the upper
field amplitudes, are equal. For $\Theta=\pi$, the $\diamondsuit$
system imitates a V-configuration (a). Mapping to the
$\Xi$-configuration, (b), is obtained for $\Theta=0$ and when
$|e\rangle$ is metastable. Similarly, the mapping to the
$\Lambda$-configuration (c) is obtained for $\Theta=0$ and
metastable intermediate states.} \label{Fig:Schemes} \end{center}
\end{figure*}

Before entering the detailed discussion of the numerical results,
it is instructive to review some basic properties of the
$\diamondsuit$--level scheme, which significantly affect its
response to light-propagation. Special symmetries of this
configuration are encountered when the laser amplitudes,
resonantly driving the upper (lower) transitions, are initially
equal, namely when \begin{eqnarray} \label{Condition:1}
&&\mathcal{G}_{e1}=\mathcal{G}_{e2}=\mathcal{G}_e\\
&&\mathcal{G}_{1g}=\mathcal{G}_{2g}=\mathcal{G}_g.
\label{Condition:2} \end{eqnarray} In this regime, the
Hamiltonian~(\ref{Ham:phase}) substantially simplifies for some
values of the phase. In particular, for $$\Theta=0,\pi$$ (modulus
$2\pi$) the dynamics can be mapped to those of well-known
three--level schemes~\cite{Morigi02}.  Insight is gained by
studying Hamiltonian~(\ref{Ham:phase}) in a convenient orthogonal
basis set of atomic states.

For $\Theta=\pi$, Hamiltonian~(\ref{Ham:phase}) describes coherent
coupling within the orthogonal subspaces
$\{\ket{g},\ket{\Psi_{12}^+}\}$ and 
$\{\ket{e},\ket{\Psi_{12}^-}\}$, where
$\ket{\Psi_{12}^{\pm}}=\left[\ket{1}\pm\ket{2}\right]/\sqrt{2}$.
These subspaces are decoupled: state $\ket{\Psi_{12}^+}$ is
decoupled from $|e\rangle$ and $\ket{\Psi_{12}^-}$ from
$|g\rangle$ by destructive interference. Spontaneous decay
eventually pumps the atom into the subspace
$\{\ket{g},\ket{\Psi_{12}^+}\}$, see Fig.~\ref{Fig:Schemes}(a).
Hence, in the stationary regime the excited state is depleted and
the atomic levels which scatter light can be mapped onto a V-level
scheme.

For $\Theta=0$ and under
condition~(\ref{Condition:1}),~(\ref{Condition:2}), the
Hamiltonian~(\ref{Ham:phase}) can be written in two equivalent
ways: It can describe coherent scattering among the orthogonal
levels $\{|1\rangle,|\Psi_{eg}^+\rangle,|2\rangle\}$, forming a
$\Lambda$-configuration, while state $\ket{\Psi_{eg}^-}$ is
decoupled, or alternatively, it can describe coherent scattering
among the orthogonal levels
$\{|g\rangle,|\Psi_{12}^+\rangle,|e\rangle\}$, forming a
$\Xi$-level scheme, while state $\ket{\Psi_{12}^-}$ is decoupled.
Here, $\ket{\Psi_{eg}^{\pm}}$ are symmetric and antisymmetric superpositions of the states
$|e\rangle$ and $|g\rangle$. However, if spontaneous decay is
included, the two schemes are not equivalent. The relaxation
processes select one configuration over the other depending on the
stability of state $\ket{\Psi_{eg}^{-}}$, which decays at a rate
$\gamma_e$, with respect to the stability of state
$\ket{\Psi_{12}^-}$, which decays at a rate
$\gamma_{1g}+\gamma_{2g}$. It is then important to introduce the
parameter \begin{equation} \label{Alpha}
\alpha=\frac{\gamma_e}{\gamma_{1g}+\gamma_{2g}} \end{equation}
which is the ratio between the decay rates of the two decoupled
states, or, equivalently, the ratio between the decay of the
excited and intermediate states. Hence, for $\alpha\ll 1$ (i.e.
the excited state is longer lived than the intermediate ones),
$\ket{\Psi_{12}^-}$ is essentially empty, and the effective
dynamics can be mapped to a $\Xi$-level scheme, see
Fig.~\ref{Fig:Schemes}(b). For $\alpha\gg 1$ instead (i.e. the
intermediate state is longer lived than the excited one),
$\ket{\Psi_{eg}^-}$ is empty, and the effective dynamics can be
mapped to a $\Lambda$-level scheme, see Fig.~\ref{Fig:Schemes}(c).
Hence, if the ratio $\alpha$ is sufficiently different from unity,
the dynamics of the $\diamondsuit$-scheme can be mapped to
three-level schemes and leads to coherent population trapping
(CPT)~\cite{EIT} in the stationary state. In the $\Lambda$ case
($\alpha\gg 1$), a large coherence between the intermediate states
is observed as reported in the transient dynamics of pulse
propagation in a medium of
$\diamondsuit$-atoms~\cite{VandenHeuwell}. Here, for some
parameter regimes one can observe population inversion at steady
state on the transition
$\ket{g}\to\ket{1},\ket{2}$~\cite{Morigi02}. In the $\Xi$ case
($\alpha\ll 1$), a macroscopic coherence between ground and
excited states is created. For some parameter regimes one can
observe population inversion at steady state on the transition
$\ket{1},\ket{2}\to\ket{e}$~\cite{Stroud76}.

These properties have important consequences for the propagation
dynamics. We note that for $\Theta=0,\pi$ the components of the
polarizations ${\rm Re}(\rho_{1g})$, ${\rm Re}(\rho_{2g})$, ${\rm
Re}(\rho_{1e})$, ${\rm Re}(\rho_{2e} \exp ({\rm i} \Theta))$
vanish. This means that the field phases remain constant upon
propagation in agreement with
Eqs.~(\ref{phipropa:g}),~(\ref{phipropa:e1})
and~(\ref{phipropa:e2}). Hence, if at the input \begin{equation}
\label{parity} \Theta(\xi=0)=0,\,\pi, \end{equation} then
\begin{equation} \label{Theta:0-pi}
\frac{\partial\Theta}{\partial\xi}=0 \end{equation} and the
relative phase remains constant during propagation along the
medium. We recall that for $\Theta=\pi$ we observe a V-type dynamics
(from now on denoted as destructive interference) and for $\Theta=0$
metastable CPT on a $\Xi$ or $\Lambda$-scheme (from now on denoted
as constructive interference). Hence, from these simple
considerations we expect that for different values of the input
phase and relaxation rates, energy will be dissipated at very
different rates along the medium.

\subsection{Destructive interference in the atomic excitations}
\label{Sec:Pi}

For $\Theta(0)=\pi$ the atoms are perfectly decoupled from the
upper fields independently of their intensity and the upper state
is empty as described in \ref{Sec:Symmetries}. Destructive
interference makes the polarizations of the transitions between
the intermediate and the upper states as well as the population of
the excited state to vanish identically, i.e.,
$\rho_{e1}=\rho_{e2}=\rho_{ee}=0$~\cite{Morigi02}.
Correspondingly, the dynamics of light propagation of the lower
fields is expected to be that encountered in a medium of V-atoms.

\begin{figure}[h!] \begin{center}
\includegraphics[width=6.5cm]{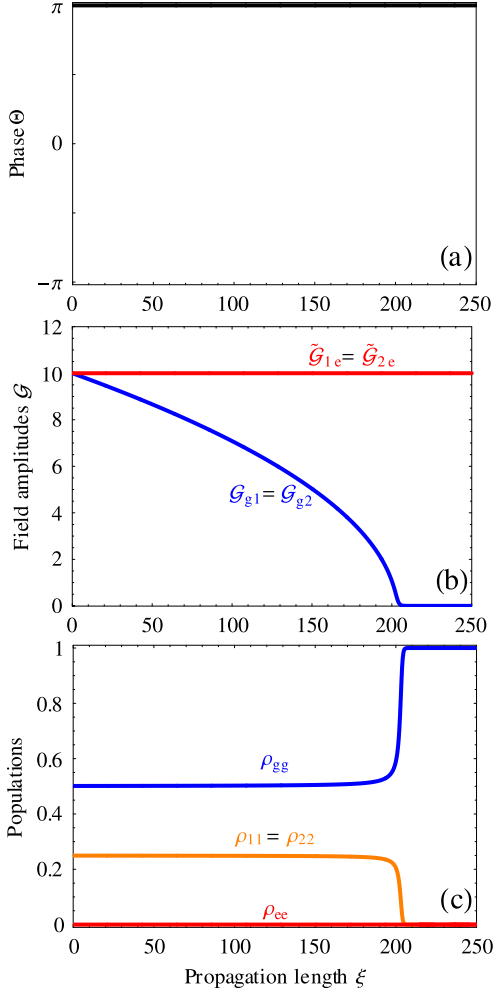} 
\caption{(color online) Propagation of the phase (a), field amplitudes (b) and the corresponding atomic state populations (c) for an input phase of $\Theta(0)=\pi$. Here, $\tilde{\mathcal{G}}_e$ is the rescaled amplitude as in Eq.~(\ref{G_e}). The phase is constant and the upper fields propagate unperturbed through the medium, while the excited state $|e\rangle$ remains depleted. The behavior is independent of $\mathcal{G}_e$ and $\gamma_e$ and thus of $\alpha$.}
\label{piprop} \end{center} \end{figure}

Figure~\ref{piprop}(a) displays the propagation dynamics along the
medium for $\Theta(0)=\pi$ and equal initial field amplitudes,
$\mathcal{G}_{ij}(0)=\mathcal{G}_0$. Here, one sees that the upper
fields propagate through the medium as if it were transparent,
keeping a constant value. The amplitudes of the lower fields
display identical decays. Figure~\ref{piprop}(b) presents the
corresponding populations of the energy levels along the medium.
The energy level $|e\rangle$ remains depleted while the
intermediate states $|1\rangle$ and $|2\rangle$ maintain the same
population as a function of $\xi$ corresponding to the fact that the lower
fields decay identically along the medium. The value of ground and
intermediate state populations is the saturation value of the
corresponding dipole transition until about $\xi\sim 200$ when the
lower fields $\mathcal{G}_{jg}(\xi)$ do not saturate the
transition any longer. After this penetration length only the
ground state is appreciably occupied. Note that these dynamics are
independent of the upper field amplitudes, as they remain
decoupled from the atoms.

We can find an analytic expression for the dynamics shown in
Fig.~\ref{piprop} and for the propagation length of the lower
fields by solving the propagation equations~(\ref{Ham:phase}) and
(\ref{epropa:g})-(\ref{phipropa:e2}) for $\Theta(0)=\Theta=\pi$.
Setting $\mathcal{G}_{ej}=\mathcal{G}_e$ and
$\mathcal{G}_{jg}=\mathcal{G}_g$, we obtain the equations for the
dimensionless amplitudes \begin{align} \frac{\partial
\mathcal{G}_g}{\partial \xi}&=-\frac{\mathcal{G}_g}
{1+4\mathcal{G}_g^2}, \label{pi1}\\
\frac{\partial \mathcal{G}_e}{\partial \xi}&=0.\label{pi:12}
\end{align} Here the right hand side in Eq.~(\ref{pi:12}) vanishes
since ${\rm Im}\{\rho_{ij}(\Theta=\pi)\}=0$. Therefore, the
relative phase $\Theta$ and the upper field amplitudes
$\mathcal{G}_e$ are constant along the medium and the medium is
transparent for the upper fields. Equation~(\ref{pi1}) is the
equation for an electric field propagating in a medium of resonant
dipoles so that the lower field amplitudes $\mathcal{G}_g$ decay
during propagation at a rate that depends only on the value of
$\mathcal{G}_g$ itself. In the case of large input intensities
(see Fig.~\ref{piprop}) a simple equation for $\mathcal{G}_g(\xi)$
is obtained~\cite{AllenEberly} \begin{equation} \mathcal{G}_g(\xi)
= \sqrt{\mathcal{G}_g^2(0)-\xi/2} \end{equation} allowing for an
estimate of the penetration depth ($2 \mathcal{G}_g^2(0)$) of the
lower fields in the medium.

\subsection{Constructive interference in the atomic excitations}
\label{Sec:Zero}

\begin{figure}[ht] \includegraphics[width=6.5cm]{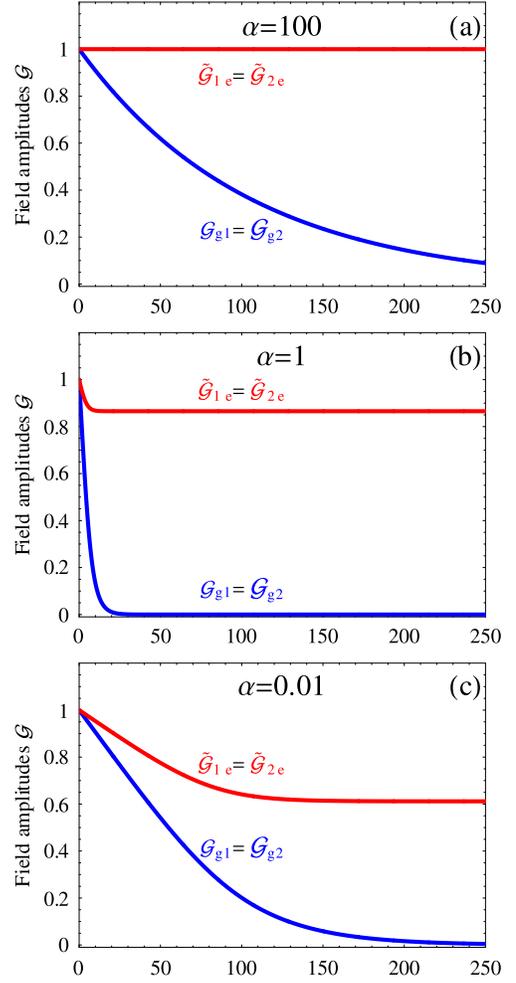}
\caption{(color online) Field amplitudes as a function of the propagation length $\xi$ for input parameters $\Theta=0$, $\mathcal{\tilde{G}}_e=\mathcal{G}_g=1$ and different ratios between the decay rates: long lived intermediate states, $\alpha=100$, in (a), balanced decay rates, $\alpha=1$, in (b), and long lived excited state, $\alpha=0.01$, shown in (c). Correspondingly, the phase $\Theta=0$ remains constant along the medium (not shown). The rate of dissipation is critically determined by $\alpha$ and is slower for $\alpha$ sufficiently larger or smaller
than unity. } \label{Fig:3}
\end{figure}

\begin{figure}[ht] \includegraphics[width=6.5cm]{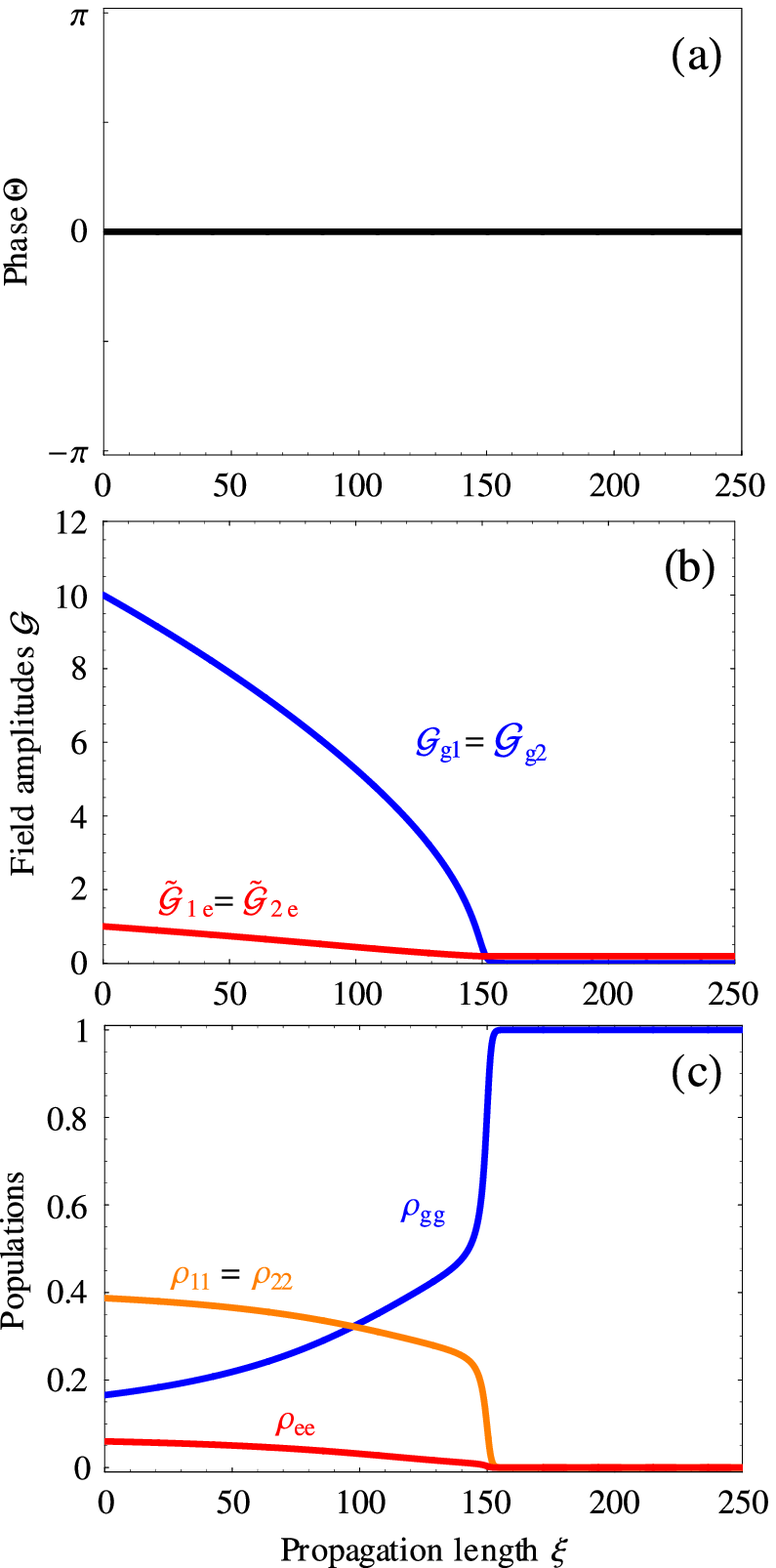}
\caption{(color online) Propagation of phase (a), field amplitudes (b) and associated populations (c) for input parameters
$\Theta=0$, $\mathcal{\tilde{G}}_e=1$, $\mathcal{G}_g=10$, and $\alpha=10$. Population inversion between the intermediate
levels and the ground level is found along the medium until $\xi \sim 100$. } \label{Fig:4}
\end{figure}

As discussed in section \ref{Sec:Symmetries}, for $\Theta=0$ the
response of the system is similar to that of a $\Xi$ or of a
$\Lambda$ level scheme, depending on the ratio of the decay rates
$\alpha$ in Eq.~(\ref{Alpha}). Atomic coherences between either
the intermediate states or the ground and excited state may form,
and correspondingly the imaginary part of the polarizations may
become very small, thus reducing dissipation.

Figure~\ref{Fig:3} displays the propagation dynamics along the
medium for different values of the ratio $\alpha$ for
$\Theta(0)=0$. For $\alpha\gg 1$ and $\alpha\ll 1$ the amplitudes
decay slowly as a function of $\xi$, as expected from the
formation of EIT-coherences. Figure~\ref{Fig:4} shows light
propagation for the case when the initial conditions of the fields
give rise to population inversion at steady state due to
metastable CPT. We find that population inversion is maintained
until $\xi \sim 100$ along the absorbing medium, but it
gradually decreases, since the atomic coherences that are
supporting CPT are not stable.

In the simulations of Figures~\ref{Fig:3} and~\ref{Fig:4}, the
lower (upper) field amplitudes remain equal during propagation. If
we assume that $\mathcal{G}_{ej}=\mathcal{G}_e$ and
$\mathcal{G}_{jg}=\mathcal{G}_g$ for all relevant $\xi$, then the
propagation equations for the amplitudes reduce to \begin{align}
\label{2pi:1} \frac{\partial \mathcal{G}_e}{\partial \xi}&=
-\frac{\mathcal{G}_e \; \mathcal{G}_g^2 \; \alpha}{D_0}
\; (1+2\alpha),\\
\label{2pi:12} \frac{\partial \mathcal{G}_g}{\partial \xi}&=
-\frac{\mathcal{G}_g \; \alpha}{D_0} \; \left [ \mathcal{G}_g^2 +\alpha
+2\alpha^2 +\mathcal{G}_e^2 \; (1 +\alpha) \right ],
\end{align}
with
\begin{eqnarray}
D_0 &=&\mathcal{G}_e^4 \; (1+\alpha) +(1+4 \mathcal{G}_g^2) \;
\alpha \;
(\mathcal{G}_g^2 +\alpha +2\alpha^2) \nonumber \\
& &+\mathcal{G}_e^2 \; \left [ \alpha \;(2+ 3\alpha)
+\mathcal{G}_g^2 \; (1+ 3\alpha +2\alpha^2) \right ] \, .
\end{eqnarray} Equations~(\ref{2pi:1}) and~(\ref{2pi:12}) describe
the dissipative propagation of the field amplitudes, and exhibit a
nonlinear dependence on the amplitudes and the ratio $\alpha$ of
the decay constants. Here, one can see that for different values
of $\alpha$ the absorption lengths can vary by orders of
magnitude. Limiting cases are found for $\alpha\to 0$, i.e. when
the excited state is stable, and for $\alpha\to\infty$, i.e., when
the intermediate states are stable. In these cases, the right hand
sides of Eqs.~(\ref{2pi:1}) and~(\ref{2pi:12}) vanish, damping is
absent, and light propagates through the medium as if it were
transparent~\cite{Footnote}.

\begin{figure*}[ht] \includegraphics[width=13.5cm]{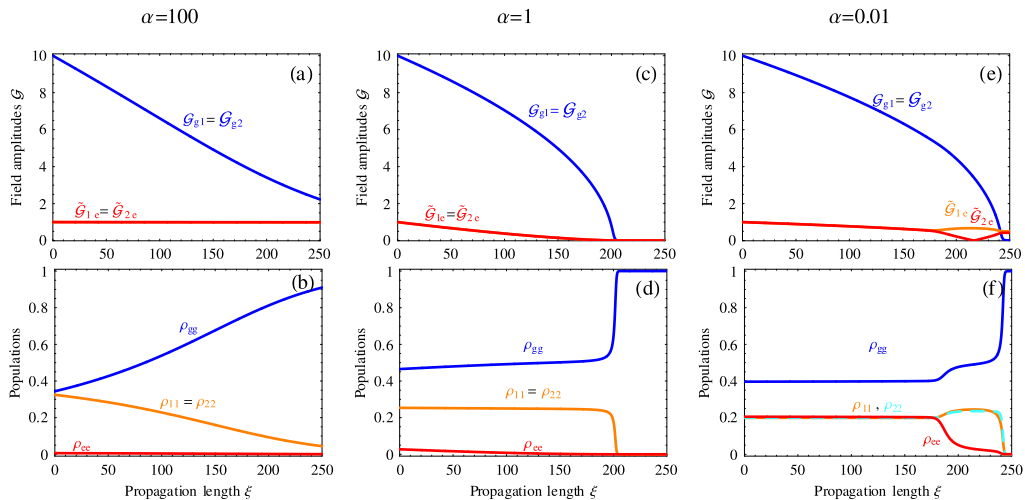}
\caption{(color online) Propagation of fields (upper row) and
corresponding atomic states populations (lower row) as a function
of the propagation length and for different values of the excited
state decay rate. The initial conditions are $\Theta(0)=0$,
$\mathcal{G}_g(0)=10$ and $\tilde{\mathcal{G}}_e(0)=1$. Here, (a)
and (b) correspond to the case $\alpha=100$, (c) and (d) to the
case $\alpha=1$, and (e) and (f) to the case $\alpha=0.01$. In
this latter case, the upper field amplitudes,
$\tilde{\mathcal{G}}_{e1}$ and $\tilde{\mathcal{G}}_{e2}$, become
different: energy is transferred from  one field to the other,
such that the amplitude of one increases while the other gradually
vanishes. At this point, the phase $\Theta$ jumps to the value
$\pi$, and energy is redistributed between the upper fields till
they reach the same value. This behavior hints to an instability
of the phase value $\Theta=0$, which seems to be triggered by
numerical fluctuations of the values of the upper field
amplitudes. } \label{thetanullprop:1} \end{figure*}

\subsection{Stability of interference under amplitude
fluctuations} \label{Sec:Stable}

\begin{figure}[h] \includegraphics[width=6.5cm]{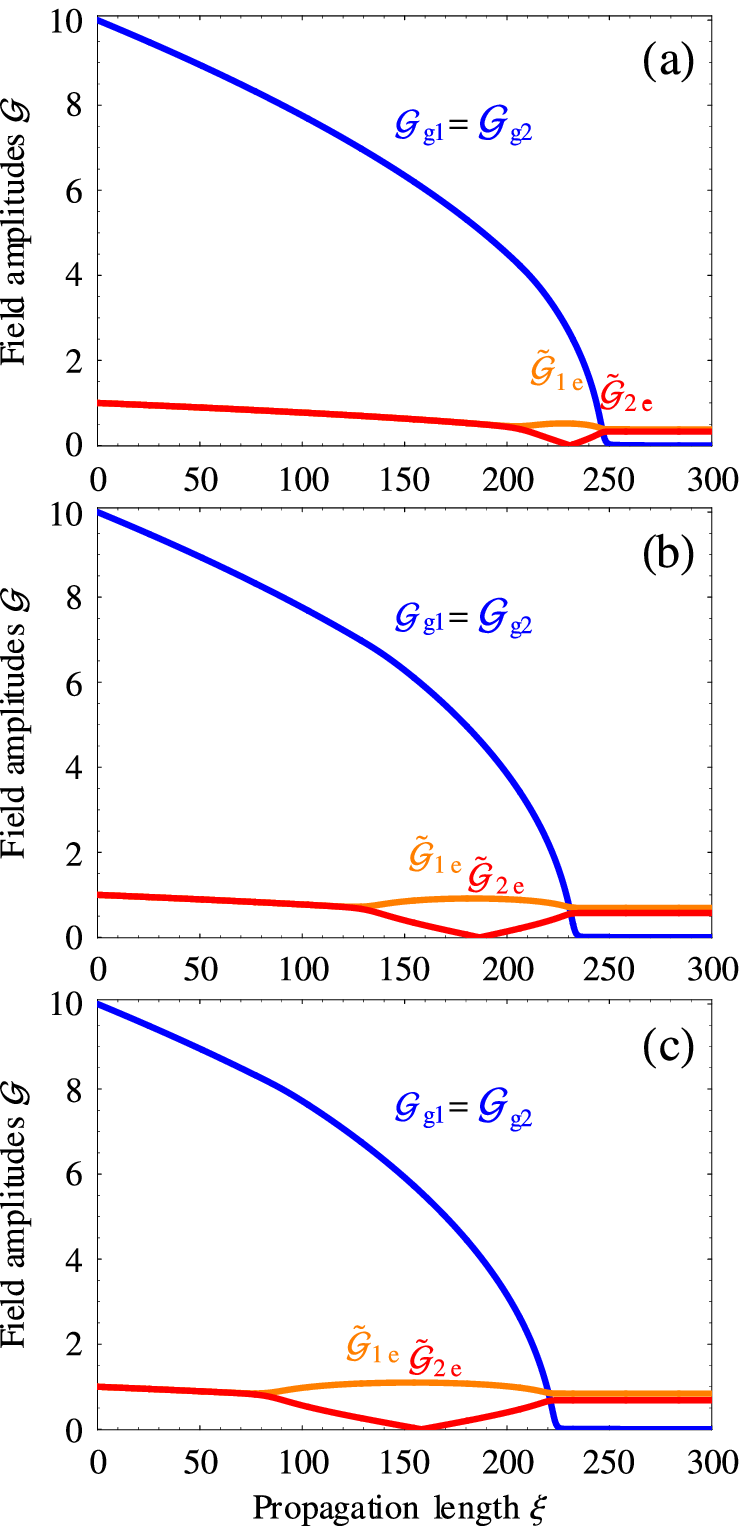}
\caption{(color online) Study of the instability, shown in
Fig.~\ref{thetanullprop:1}(e), as a function of different size of
the upper field amplitude fluctuation $\delta\mathcal{G}_{e}$. The
curves show the field amplitudes as a function or the propagation length $\xi$, for
$\alpha=0.01$, $\mathcal{G}_{jg}(0)=\mathcal{G}_{g}=10$,
$\mathcal{\tilde{G}}_{e1}(0)=1$ and
$\mathcal{\tilde{G}}_{e2}(0)=1-\delta\mathcal{G}_{e}$ with (a)
$\delta\mathcal{G}_{e}=10^{-16}$, (b)
$\delta\mathcal{G}_{e}=10^{-12}$ and (c)
$\delta\mathcal{G}_{e}=10^{-8}$.} \label{thetanullprop:2}
\end{figure}

In the cases discussed so far, the input phase is a constant of
the propagation, and the amplitudes of the upper fields, as well
as the amplitudes of the lower fields remain equal along the
medium. We now address the question of stability of these
configurations against phase and amplitude fluctuations.

Numerical investigations show that at $\Theta=\pi$ the V-level
dynamics is robust against phase and amplitude fluctuations, from
which we infer that this is a stable configuration. It should be
remarked that the overall behavior is transient in that the medium
dissipates the lower fields until (well inside the medium) the
atoms are all found in the ground state.

The case $\Theta=0$ is more peculiar. In the cases discussed in
Sec.~\ref{Sec:Zero}, energy is exchanged between upper and lower
fields until the lower field amplitudes go below saturation. Then,
the upper fields decouple as the population of the intermediate
states becomes negligible. In order to study the long term
dynamics of propagation, we now focus on the regime where the
lower transitions are driven well above saturation and where we
may expect different length scales for the propagation of the
upper and lower fields.

Figure~\ref{thetanullprop:1} displays propagation when the lower
transitions are driven well above saturation for different values
of $\alpha$. The dynamics observed in the $\alpha=1$ case
separates the regimes corresponding to a $\Xi$-like response for
$\alpha \ll 1$, and a $\Lambda$-like response for $\alpha \gg 1$.
For the separating case of $\alpha=1$ in
Fig.~\ref{thetanullprop:1}(c) and (d) we find that the damping of
the lower fields below saturation is accompanied by a drop of the
population from intermediate to ground states. For $\alpha=100$,
shown in Fig.~\ref{thetanullprop:1}(a) and (b) propagation is
characterized by EIT-like coherences between the intermediate
states, which are established through the medium by the action of
the lower fields. These coherences increase the penetration depth
of the lower fields in the medium and allow the upper fields to
propagate quasi undamped. This is consistent with the behaviour
discussed in Sec.~\ref{Sec:Zero}. However, for $\alpha=0.01$ in
Fig.~\ref{thetanullprop:1}(d) and~(e) we observe a clear deviation
from the symmetric decay of the upper field amplitudes at long
propagation length.

We now focus on this case, which exhibits novel features with
respect to the cases studied so far. In
Fig.~\ref{thetanullprop:1}(e) one observes that the upper field
amplitudes, $\mathcal{G}_{e1}$ and $\mathcal{G}_{e2}$, initially
equal to each other, undergo a transient behavior where they
become different: energy is transferred from  one field to the
other, such that the amplitude of one increases while the other
gradually vanishes. This behavior is accompanied by a depletion of
the excited state, while the intermediate states continue to be
equally populated. At the same time of the vanishing of one of the
upper field amplitudes, the phase $\Theta$ jumps from $0$ to $\pi$
and energy is redistributed between the two upper fields till they
reach almost the same value. After this transient, the field
amplitudes of the excited states remain at a constant value across
the medium. Correspondingly, during and after this transient, the
excited state population in Fig.~\ref{thetanullprop:1}(f)
decreases until it reaches zero. This remarkable behavior hints to
an instability of the phase value $\Theta=0$, which seems to be
triggered here by numerical fluctuations of the values of the
upper field amplitudes. Such conjecture is supported by the
numerical analysis shown in Fig.~\ref{thetanullprop:2}, where we
have introduced fluctuations between the initial values of the
upper field amplitudes. As the initial discrepancy increases, the
splitting of the upper field amplitudes appears at earlier
locations in the medium although the behavior of the lower fields
is unaffected. More detailed investigations on populations and
phases show that the imbalance between the upper field amplitudes
induces a depletion of the excited state until the vanishing of
one of the upper amplitudes forces a phase jump to the value
$\Theta=\pi$ and the upper fields decouple from the atom. After
the phase jump, the upper field amplitudes tend to recover an
equal value, but they decouple from the atoms once the lower field
amplitudes have vanished.

An explanation of the phase jump from $\Theta=0$ to $\Theta=\pi$
is the tendency of the system to minimize the rate of dissipation
in a way reminiscent of what is observed in Four-Wave Mixing
experiments where interference effects are generated in order to
minimize spontaneous emission~\cite{Boyd}. We also note that with
increasing values of $\alpha$, the splitting of the upper fields
for the same initial difference in their amplitudes is
progressively delayed inside the medium and eventually disappears.

\subsection{Generic phase at the input fields}

\begin{figure*}[htb]
\begin{center}
\includegraphics[width=12cm]{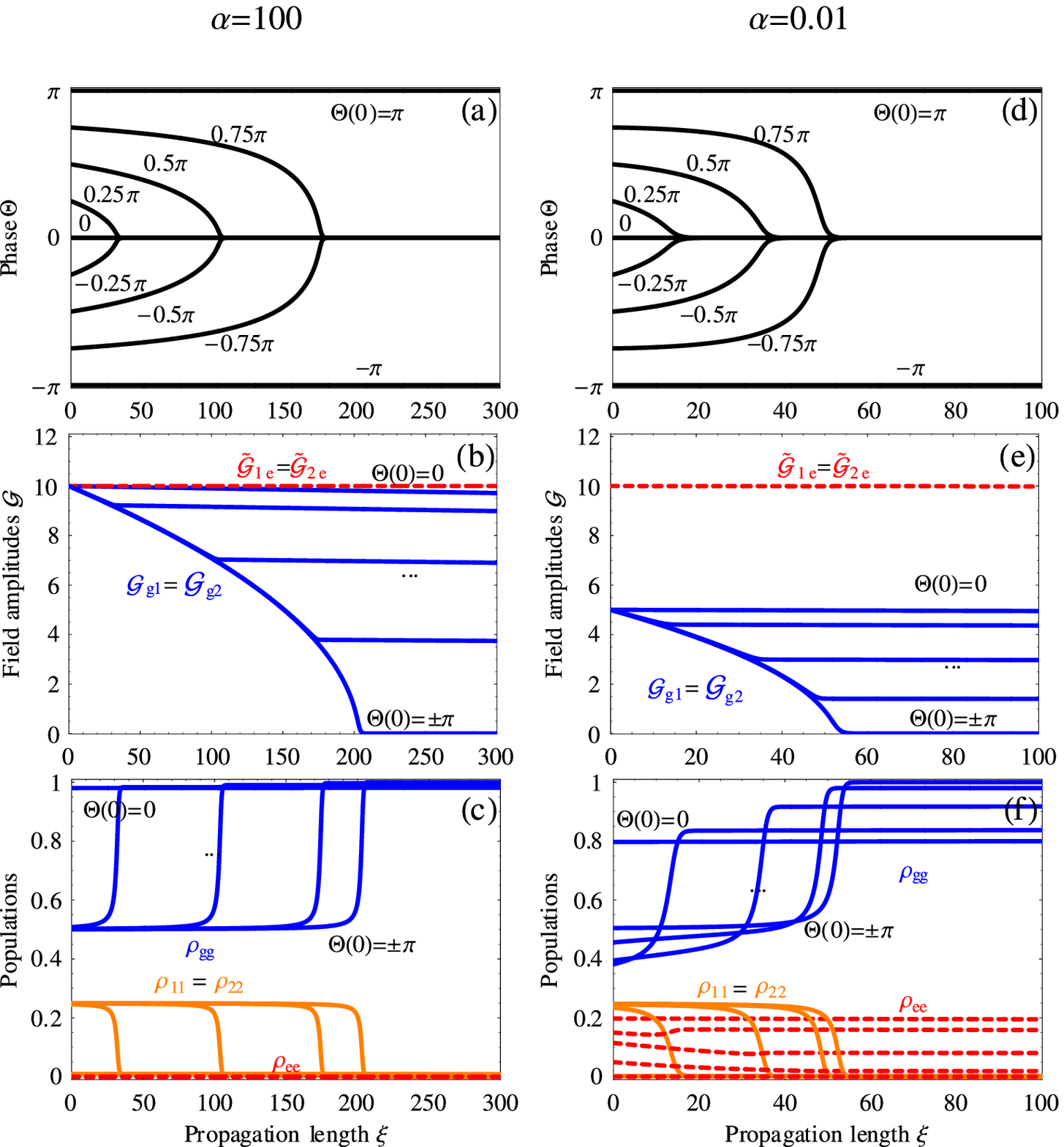} \caption{(color online) Propagation of relative phase (a), field amplitudes (b), and populations of the atomic levels (c), for various input phases $\Theta(0)=0, \pm \pi/4, \pm \pi/2, \pm 3\pi/4, \pi$. The upper fields are saturated at least as deeply as the lower fields. Curves (a-c) are obtained for input parameters $\alpha=100$, and $\mathcal{G}_g=\mathcal{\tilde{G}}_e=10$. Curves (d-f) have input parameters $\alpha=0.01$, $\mathcal{G}_g=5$ and $\mathcal{\tilde{G}}_e=10$. During propagation the phase tends to $\Theta=0$. Once this phase is reached, the rate of energy dissipation along the medium changes abruptly to a significantly lower level. In the case(d-f) this change is accompanied by the establishing of population inversion on the transition $\ket{1},\ket{2}\to\ket{e}$, see (f), due to the formation of EIT-coherences.} \label{Fig:Phase:0}
\end{center} \end{figure*}

Having identified and investigated two special values of the input phase, we now address the question, how the phase evolves starting from a generic input value, and correspondingly, how light propagates and is dissipated along the medium. We restrict to the configuration with initially equal lower field amplitudes and equal upper field amplitudes~(\ref{Condition:1}) and~(\ref{Condition:2}), and vary the input phase $\Theta(0)=\pi/2$ in steps of $\pi/4$.

\begin{figure*}[htb]
\begin{center}
\includegraphics[width=12cm]{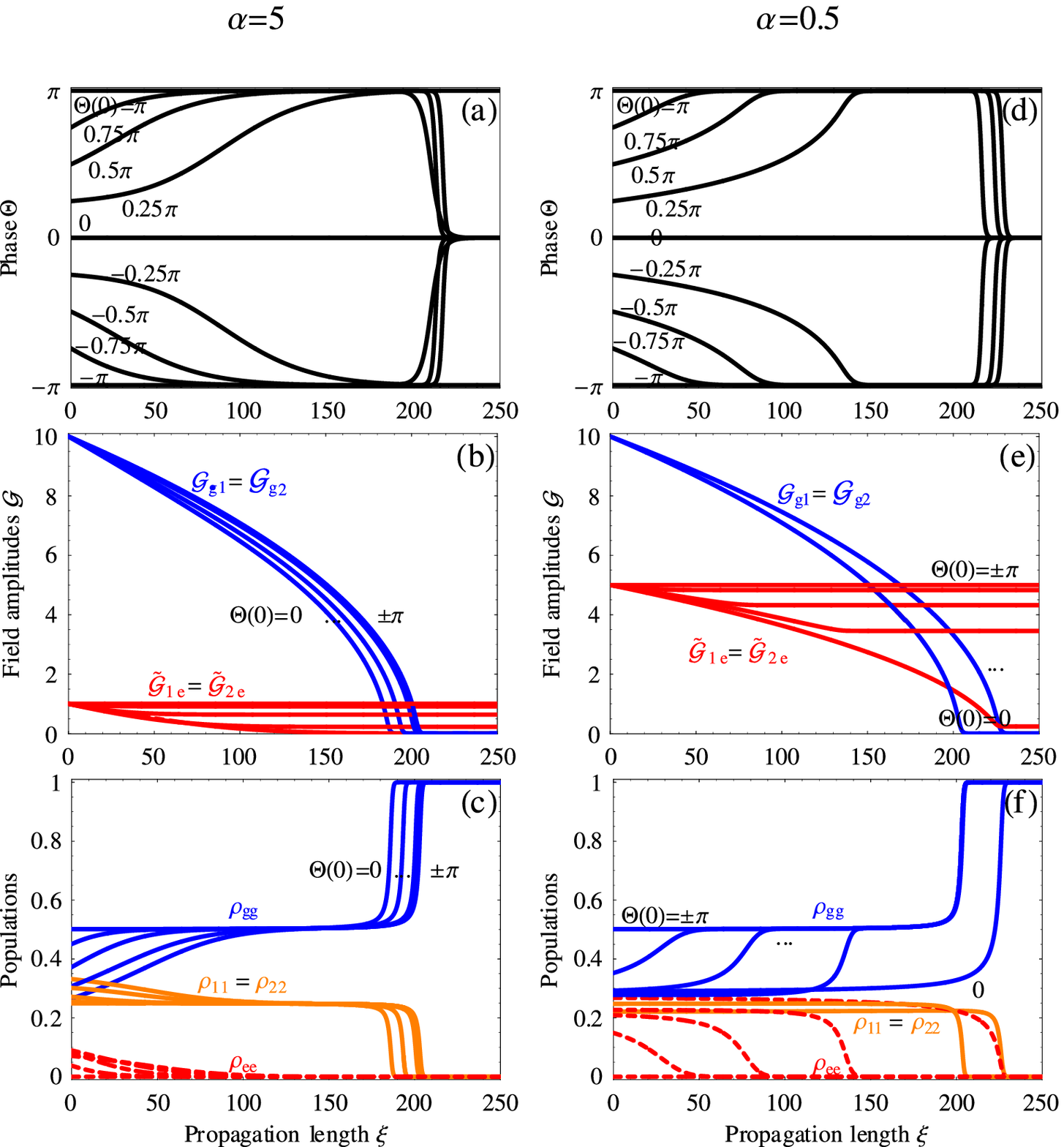}
\caption{(color online) Propagation of relative phase (a), field amplitudes (b), and populations of the atomic levels (c),
for various input phases $\Theta(0)=0, \pm \pi/4, \pm \pi/2, \pm 3\pi/4, \pi$.  Curves (a-c) are obtained for input
parameters $\alpha=5$, and $\mathcal{G}_g=10$, $\mathcal{\tilde{G}}_e=1$, and curves (d-f) for $\alpha=0.5$ and the same
initial amplitudes.  Here, the lower transitions are driven well above saturation and during a transient regime the phase
tends to the value $\pi$, while the upper fields decouple. The transition to the value $\Theta=0$
at large values of $\xi$ is an artifact, since for these lengths the
lower fields are very weak and the atoms are essentially in the ground
state.} \label{Fig:Phase:pi} \end{center} \end{figure*}

Figures~\ref{Fig:Phase:0} and~\ref{Fig:Phase:pi} display the amplitude and the relative phase of the propagating fields for different values of the excited state decay rate: $\alpha=100$ and $\alpha=0.01$. Although the lower field amplitudes are clearly damped for all values of $\alpha$, the mechanism of radiation dissipation depends on $\alpha$ and on the initial strengths of the field amplitudes. This can be associated with a particular evolution of the phase along the medium, which in the cases displayed in Fig.~\ref{Fig:Phase:0} reaches the stable value $\Theta=0$, and in the cases displayed in Fig.~\ref{Fig:Phase:pi} tends first
to the value $\pi$ before eventually reaching $\Theta=0$. The choice between these behaviors depends on the input amplitudes of the fields.  We now discuss these two behaviors in detail.

In Fig.~\ref{Fig:Phase:0} all atomic transitions are driven at
saturation, and the saturation of the upper transitions is larger
or equal to that of the lower transitions. We observe that the
relative phase of the fields tends to the zero value. Before this
value is reached, radiation is damped at a fast rate. Once
$\Theta=0$, the rate of damping of the lower field amplitudes
changes abruptly to a lower value. This sudden change occurs at a
propagation length determined by the typical absorption length of
the fast decaying transition. The system simulates a
V-configuration, thereby switching to an EIT-like response. A
similar kind of behavior is also observed in a medium of the
Double-$\Lambda$ atoms where EIT-coherences are established
between the two stable states~\cite{Korsunsky99}. In the
$\diamondsuit$ configuration the coherences and interferences are
transient~\cite{Windholz96}. For a slower decay of the excited
state, however, the system can also switch to a $\Xi$-dynamics and
exhibit a transient coherence between ground and excited states. A
manifestation of this phenomenon is population inversion between
the excited and the intermediate states along the medium in
Fig.~\ref{Fig:Phase:0}(f).

In Fig.~\ref{Fig:Phase:pi} the lower transitions are driven well
above saturation, and the corresponding saturation parameter is
larger than the saturation parameter of the upper fields. Here,
during a transient regime the phase slowly tends to $\Theta=\pi$.
Nonetheless, the tendency of the medium for long propagation
lengths is to eventually decouple upper fields and atoms, i.e. to
switch to a V-dynamics. The onset of this dynamics depends
critically on the value of $\mathcal{G}_g$, which must well
saturate the lower transitions with respect to $\mathcal{G}_e$ in
order to populate the intermediate states on a time scale shorter
than their decay rate, but long enough for incoherent decay of the
upper state to take place. This behavior is in agreement with
Sec.~\ref{Sec:Stable}, showing that when the lower transitions are
driven well above saturation the value $\Theta=\pi$ is the only
stable phase under amplitude and phase fluctuations.

\subsection{Four-wave mixing} \label{Sec:weak}

So far, we have considered input fields with equal lower and upper
field amplitudes. We now discuss propagation when one field is
initially very weak while the other three transitions are driven
at or above saturation, and study how the dynamics of energy
redistribution among the fields depends on the input parameters
and on the stability of the excited state.

\begin{figure}[h!,t,b] 
\begin{center}
\includegraphics[width=6cm]{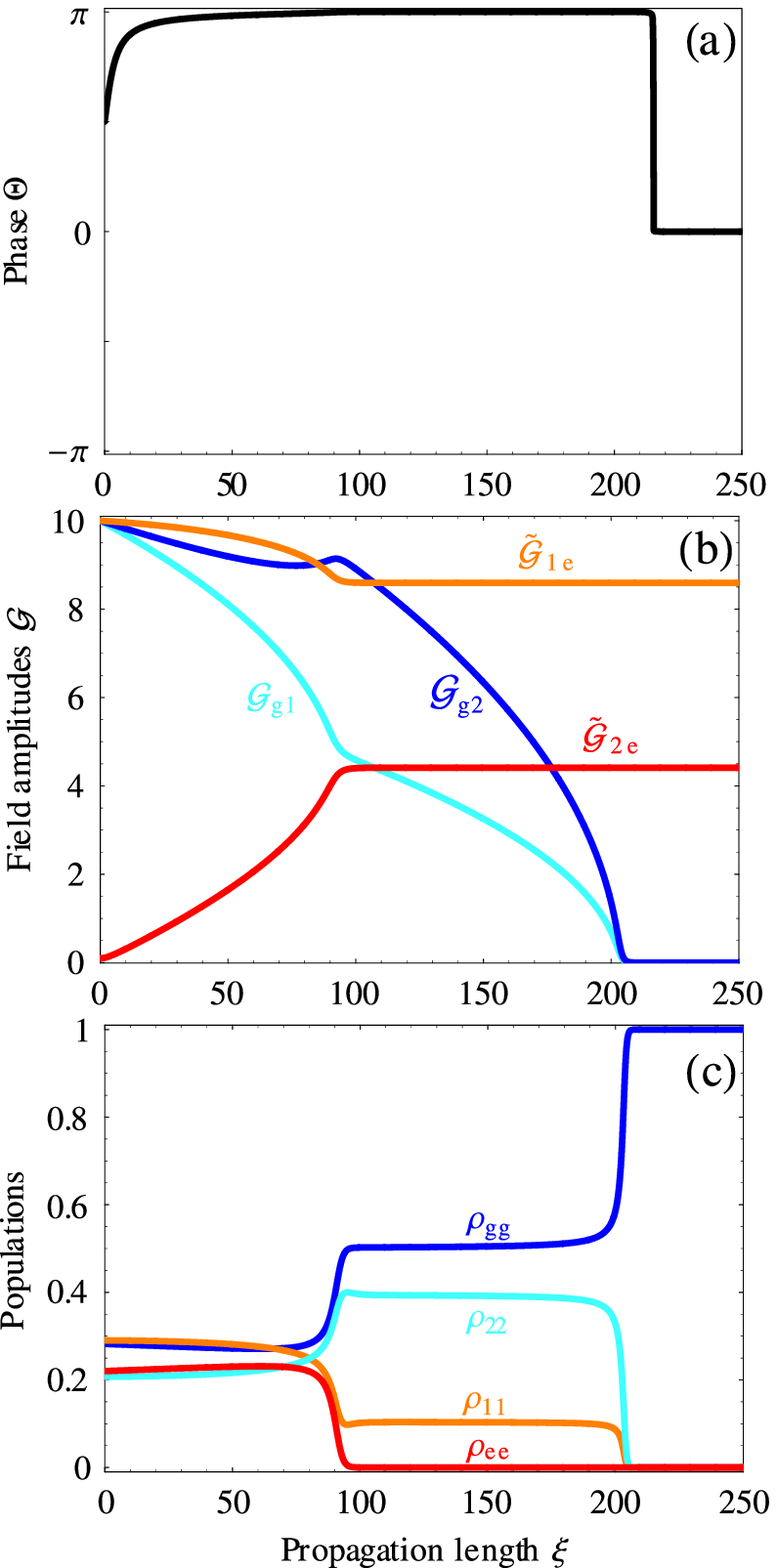} \caption{(color online) (a)
Propagation of relative phase (a), field amplitudes (b),
and atomic populations (c) for input parameters $\Theta=\pi/2$, $\alpha=0.1$,
$\mathcal{G}_{g1}=\mathcal{G}_{g2}=\mathcal{G}_{1e}=10$, and $\mathcal{G}_{2e}=0.1$. Energy is exchanged between fields
$\mathcal{G}_{2e}$ and $\mathcal{G}_{g1}$ and also between $\mathcal{G}_{1e}$ and $\mathcal{G}_{g2}$ until the excited
state decouples and the upper fields propagate freely. The jump of the phase to the value 0 is
an artifact due to all atoms being in the ground state. }
\label{Fig:4wave} \end{center} \end{figure}

\begin{figure}[h!] \begin{center}
\includegraphics[width=6cm]{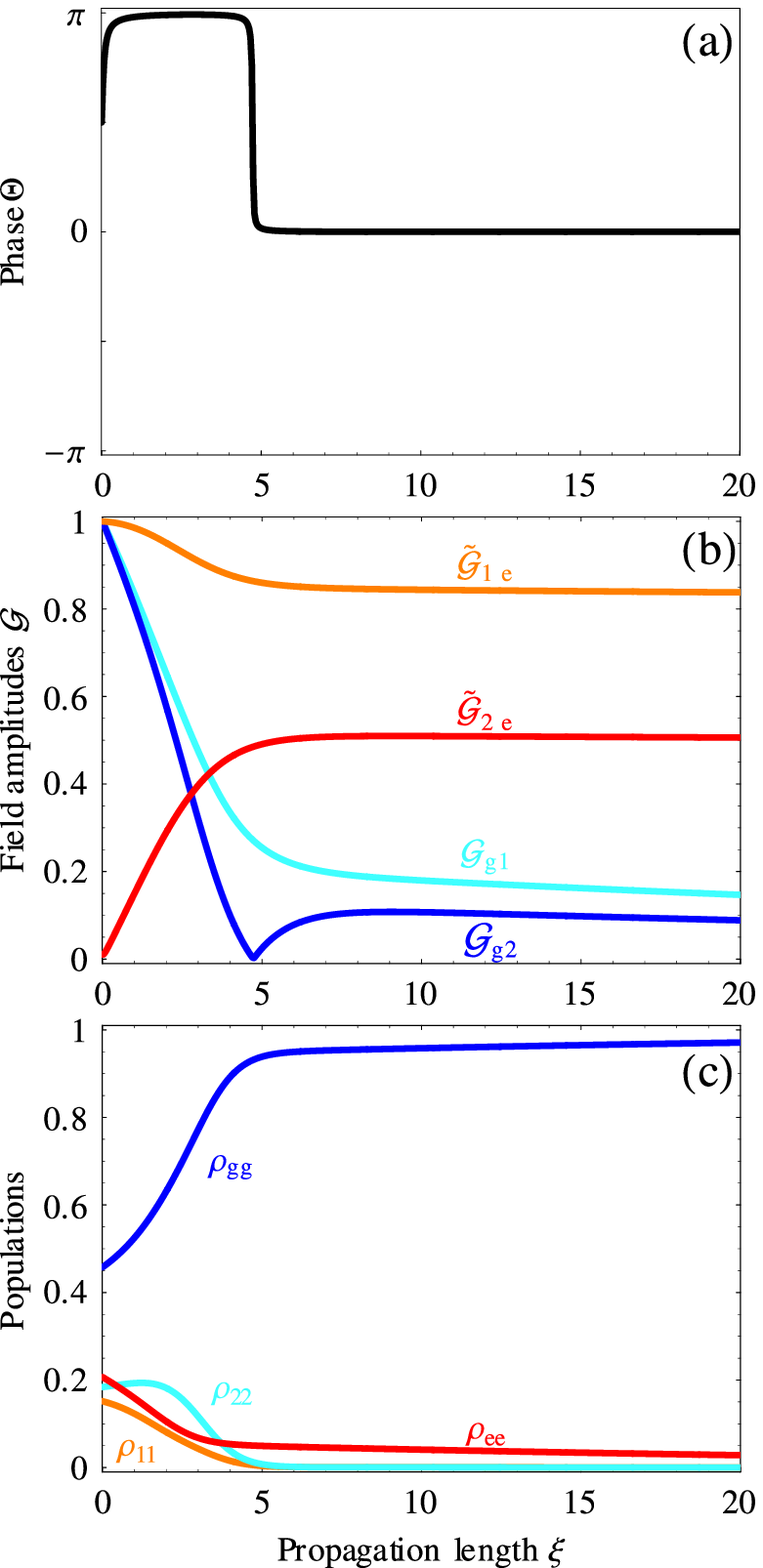} \caption{(color online) (a)
Propagation of relative phase (a), field amplitudes (b), and atomic populations (c) for input parameters $\Theta=\pi/2$,
$\alpha=0.01$, $\mathcal{G}_{g1}=\mathcal{G}_{g2}=\mathcal{\tilde{G}}_{1e}=1$, and $\mathcal{\tilde{G}}_{2e}=0.01$. The
amplitude $\mathcal{\tilde{G}}_{2e}$ while the phase approaches $\Theta=\pi$.  At the same time $\mathcal{\tilde{G}}_{1e}$
gradually vanishes at which point the phase changes to $\Theta=0$ and the fields propagate almost undamped.} \label{Fig:4wave:2}
\end{center} \end{figure}

Figures~\ref{Fig:4wave} and~\ref{Fig:4wave:2} display the field
propagation when the upper field amplitude $\mathcal{G}_{e2}$ is
very small and the phase is initially set to the value
$\Theta(0)=\pi/2$. In both figures we have assumed the excited
state to decay slower than the intermediate states, but one may
also observe amplification of the weak field under different
conditions. In Fig.~\ref{Fig:4wave} the three input fields drive
the respective transitions well above saturation. Here,
amplification of the weak field is accompanied by the asymptotic
approaching of the phase to $\Theta=\pi$. The field
$\mathcal{G}_{e2}$ is amplified until the upper state is depleted
because of interference between the upper fields. From this point
further the phase $\Theta=\pi$ is stable and the lower fields
dissipate, until they drop below saturation. The jump of the phase
to the value 0 is an artifact due to all atoms being in the ground
state.

In Fig.~\ref{Fig:4wave:2} the three input fields $\mathcal{G}_{1g},\mathcal{G}_{2g}$
and $\mathcal{G}_{e1}$ are just saturating the respective transitions. Here,
amplification of the field $\mathcal{G}_{e2}$ is accompanied by a
transient stabilization of the phase at $\Theta=\pi$. This is
accompanied by a fast decrease of the lower field amplitude
$\mathcal{G}_{g2}$, until, when $\mathcal{G}_{g2}$ vanishes, the
phase falls to $\Theta=0$. After this point the behavior changes
and $\mathcal{G}_{g2}$ first increases slightly and then decays
slowly as a function of $\xi$ in a way similar to
$\mathcal{G}_{g1}$, while the upper field amplitudes remain
constant. The final configuration supports a coherence between the
excited and the ground state, and indeed for $\Theta=0$ and this
value of $\alpha$ the dynamics can be mapped to a $\Xi$-level
scheme. In particular, due to destructive interference, the fields
are only weakly coupled to the transitions and the medium is
semitransparent. This is also supported by
Fig.~\ref{Fig:4wave:2}(c) where one sees that the population is
redistributed between the ground and the excited state while the
intermediate states are depleted. In this regime the medium is
characterized by population inversion between the excited and the
intermediate states.

\section{Discussion and conclusions} \label{Sec:Conclude}

We have investigated numerically light propagation in a medium of
atoms whose electronic levels are resonantly driven by lasers in a
$\diamondsuit$ configuration. Propagation is critically affected
by the initial parameters of the input fields and shows the
tendency to reach configurations which minimize dissipation. An
important role is played by the relative phase $\Theta$ between
the fields. It exhibits two fixed points, $\Theta=0$ and
$\Theta=\pi$, whose stability during propagation depends on the
field amplitudes and on the ratio $\alpha$ between the rates of
dissipation of excited and intermediate states. A generic input
phase evolves, in general, to one of these values, again depending
on the input amplitudes and $\alpha$.

These two metastable phase values are associated with two
different types of atomic coherences. The response of the medium,
corresponding to the phase $\Theta=0$, is characterized by the
formation of atomic coherences typical of EIT-media. Similar
behaviors have been observed for instance
in~\cite{VandenHeuwell,Windholz96} and are analogous to the
response predicted for light propagation in double-$\Lambda$
media~\cite{Korsunsky99}.

The response of the medium for the phase $\Theta=\pi$ is supported
by a different type of interference which leads to a depletion of
the upper state and to a complete decoupling of the upper fields
from the atom. For this value of the phase, the medium acts like a
$V$-level configuration. We note that this value of the phase
appears to be the preferred value for the $\diamondsuit$ medium if
the lower transitions are driven well above saturation. This
behavior is novel to our knowledge and it is reminiscent of the
phenomenon of suppression of spontaneous emission observed in
four-wave mixing studies in atomic gases~\cite{Boyd}.

In general, the system exhibits a rich dynamics and several novel
features due to atomic coherence which offer new perspectives in
control techniques in quantum electronics. These could be studied
in atomic gases where the ground state has no hyperfine multiplet,
like e.g. alkali-earth isotopes which are currently investigated
for atomic clocks~\cite{AlkaliEarth}.

In the future we will extend our analysis to the case in which the
transitions are not resonantly driven and we will address the
asymptotic behavior of the system following the lines of recent
works~\cite{Korsunsky02,Johnsson}.

\acknowledgements

The authors thank E. Arimondo, S.M. Barnett, R. Corbalan, and W.P.
Schleich for discussions and helpful comments. G.M. and S.K.-S.
acknowledge the warm hospitality of the Department of Physics at
the University of Strathclyde. This work has been partially
supported from the RTN-network CONQUEST and the scientific
Exchange Programme Germany-Spain (HA2005-0001 and D/05/50582).
G.M. is supported by the Spanish Ministerio de Educacion y
Ciencias (Ramon-y-Cajal and FIS2005-08257-C02-01). S.F-A is
supported by the Royal Society. G-L.O. thanks the CSDC of the
University of Florence (Italy) for its kind hospitality.

\begin{appendix}

\section{Macroscopic Polarization in the semiclassical limit for the
atomic motion}

We consider the dynamics of the density matrix $\varrho$ of the
atomic internal and external degrees of freedom, where the
center-of-mass degrees of freedom are treated as classical
variables. Hence, the position ${\bf x}$ and momentum ${\bf p}$
are parameters, distributed according the function $w({\bf x},{\bf
p})$ which we assume to be stationary, with $\int {\rm d}{\bf
x}{\rm d}{\bf p}w({\bf x},{\bf p})=N$ and $N$ is the number of
atoms. The spatial density of atoms $n({\bf x})$ is found from
$w({\bf x},{\bf p})$ according to $\int {\rm d}{\bf p}w({\bf
x},{\bf p})=n({\bf x})$. In this work we assume uniform density,
namely $$n({\bf x})=n$$ with $n$ constant. The master equation for
the density matrix $\varrho$, at the point $({\bf x},{\bf p})$ in
phase space has the form \begin{equation}
\dot{\varrho}=\frac{1}{{\rm i} \hbar}\left[ H({\bf x},{\bf
p};t),\varrho \right]+\mathcal{L}\varrho \end{equation} where the
Hamiltonian $H({\bf x},{\bf p};t)$ contains the coherent dynamics
of the atoms driven by the classical field, \begin{eqnarray}
H({\bf x},{\bf p};t)=\frac{{\bf p}^2}{2M}+H(z,t) \end{eqnarray}
and $H(z,t)$ is defined in Eq.~(\ref{hamilton1}). The Liouvillian
$\mathcal{L}$ in Eq.~(\ref{Liou}) describes the relaxation
processes, which we consider here to be purely radiative. The
corresponding macroscopic polarization has the form
\begin{eqnarray} {\bf P}({\bf x},t)=\int{\rm d}{\bf p}w({\bf
x},{\bf p}) \Trace\{{\bf \hat{d}}\varrho({\bf x},{\bf p})\}
\end{eqnarray} Assuming that the atomic gas has been Doppler
cooled, so that line broadening is homogeneous, the kinetic energy
can be neglected in evaluating the atomic response to the light.
By integrating over ${\bf p}$ and $x,y$ we hence obtain
Eq.~(\ref{Master:z}), whereby $\sigma(z)=\int{\rm d}{\bf p}{\rm
d}x{\rm d}y w({\bf x},{\bf p}) \sigma({\bf x},{\bf p})$, and
polarization as in Eq.~(\ref{polarisation}).

\section{Optical Bloch Equations in the {\it phase}- reference frame}

We consider Master Eq.~(\ref{master:phase}) in the reference frame of the
phase. With the notation $\tilde{\rho}_{e2}=\rho_{e2}
\;\text{exp}\left(-i\Theta\right)$, the corresponding optical Bloch
equations are given by
\begin{eqnarray}
\dot{\rho}_{ee}&=&i \frac{\Omega_{1e}}{2} (\rho_{1e} -\rho_{e1})
+i
  \frac{\Omega_{2e}}{2} (\tilde{\rho}_{2e} \label{obe1}\\
               & &-\tilde{\rho}_{e2}) -\gamma_{e} \rho_{ee}
\nonumber\\
\dot{\rho}_{11}&=&-i \frac{\Omega_{1e}}{2} (\rho_{1e} -\rho_{e1})
+i\frac{\Omega_{1g}}{2} (\rho_{g1} -\rho_{1g})\nonumber\\
               & &+\frac{\gamma_{e}}{2}\rho_{ee}
  -\gamma_{1g}\rho_{11} \\
\dot{\rho}_{22}&=&-i \frac{\Omega_{2e}}{2} (\tilde{\rho}_{2e} -
  \tilde{\rho}_{e2}) +i \frac{\Omega_{2g}}{2} (\rho_{g2}-\rho_{2g})
  \nonumber\\
               &  &+\frac{\gamma_{e}}{2}\rho_{ee} -\gamma_{2g}\rho_{22}\\
\dot{\rho}_{e1}&=&\left( i \left( \Delta_1 -\Delta_e \right)
  -\frac{\gamma_{e} +\gamma_{1g}}{2}
  \right)\rho_{e1}\\
               & &+i \frac{\Omega_{1e}}{2} (\rho_{11} -\rho_{ee}) +i
\frac{\Omega_{2e}}{2} e^{i \Theta}\rho_{21} -i
\frac{\Omega_{1g}}{2} \rho_{eg} \nonumber\\
\dot{\tilde{\rho}}_{e2}&=&\left( i \left( \Delta_2 -\Delta_e
\right)
  -\frac{\gamma_{e} +\gamma_{2g}}{2}
  \right)\tilde{\rho}_{e2}\\
                       & &+i \frac{\Omega_{2e}}{2} (\rho_{22} -\rho_{ee})
+i \frac{\Omega_{1e}}{2}e^{-i \Theta}\rho_{12} -i
\frac{\Omega_{2g}}{2} e^{-i \Theta}\rho_{eg}\nonumber
\\
\dot{\rho}_{1g}&=&-\left( i\Delta_1 +\frac{\gamma_{1g}}{2}
\right)\rho_{1g} +i
  \frac{\Omega_{1e}}{2}\rho_{eg}\\
               & &-i \frac{\Omega_{2g}}{2}\rho_{12} +i
  \frac{\Omega_{1g}}{2} (\rho_{gg} -\rho_{11})\nonumber\\
\dot{\rho}_{2g}&=&-\left( i\Delta_2 +\frac{\gamma_{2g}}{2}
\right)\rho_{2g} +i
  \frac{\Omega_{2e}}{2} e^{-i \Theta} \rho_{eg} \\
               & &-i
  \frac{\Omega_{1g}}{2}\rho_{21} +i \frac{\Omega_{2g}}{2} (\rho_{gg}
  -\rho_{22})\nonumber\\
\dot{\rho}_{12}&=&\left( i\left( \Delta_2-\Delta_1 \right)
-\frac{\gamma_{1g} +\gamma_{2g}}{2} \right)\rho_{12} +i
\frac{\Omega_{1e}}{2}e^{i \Theta} \tilde{\rho}_{e2} \nonumber\\
               & &+i\frac{\Omega_{1g}}{2}\rho_{g2} -i
  \frac{\Omega_{2g}}{2}\rho_{1g} -i  \frac{\Omega_{2e}}{2} e^{i
  \Theta} \rho_{1e}\\
\dot{\rho}_{eg}&=&-\left( i\Delta_e +\frac{\gamma_{e}}{2}
  \right)\rho_{eg} +i \frac{\Omega_{1e}}{2} \rho_{1g} \nonumber\\
               & &+i \frac{\Omega_{2e}}{2}
  e^{i \Theta} \rho_{2g} -i \frac{\Omega_{1g}}{2}\rho_{e1} -i
  \frac{\Omega_{2g}}{2}e^{i \Theta}\tilde{\rho}_{e2}.  \label{obe10}
\end{eqnarray}
where $\rho_{ij}=\rho_{ji}^*$,
$\rho_{gg}=1-\rho_{ee}-\rho_{11}-\rho_{22}$, and we have taken
$\gamma_{e1}=\gamma_{e2}=\gamma_e/2$.

\end{appendix}


\begin{thebibliography}{99}

\bibitem{EIT}
S.E. Harris, Phys. Today {\bf 50}, 36 (1997); E. Arimondo, Prog.
Opt. {\bf 35}, 259 (1996).

\bibitem{Harris04}
D.A. Braje, V. Balic, S. Goda, G.Y. Yin, and S.E. Harris,
Phys. Rev. Lett. {\bf 93}, 183601 (2004).


\bibitem{Harris05}
V. Balic, D.A. Braje, P. Kolchin, G.Y. Yin, and S.E. Harris,
Phys. Rev. Lett. {\bf 94}, 183601 (2005).


\bibitem{Lukin05}
M.D. Eisaman, L. Childress, A. Andr\'e, F. Massou, A.S. Zibrov, and M.D. Lukin,
Phys. Rev. Lett. {\bf 93}, 233602 (2004).

\bibitem{Nottelmann93}
A. Nottelmann, C. Peters, and W. Lange, Phys. Rev. Lett.
{\bf 70}, 1783 (1993)

\bibitem{Peters96}
C. Peters and W. Lange, App. Phys. B {\bf 62}, 221 (1996)


\bibitem{RIR}
J. Guo, P. R. Berman, B. Dubetsky, and G. Grynberg,
Phys. Rev. A {\bf 46}, 1426 (1992).

\bibitem{Prentiss05}
M. Vengalattore and M. Prentiss,
Phys. Rev. Lett. {\bf 95}, 243601 (2005)

\bibitem{Buckle86}
S.J. Buckle, S.M. Barnett, P.L. Knight, M.A. Lauder, and D.T.
Pegg, Optica Acta {\bf 33}, 1129 (1986).

\bibitem{Kosachiov92}
D.V. Kosachiov, B.G. Matisov, and Y.V. Rozhdestvensky, J. Phys. B:
At. Mol. Opt. Phys. {\bf 25}, 2473 (1992).

\bibitem{Korsunsky99} E.A. Korsunsky and D.V. Kosachiov, Phys.
Rev. A {\bf 60}, 4996 (1999).

\bibitem{Morigi02} G.\ Morigi, S.\ Franke-Arnold, G.-L.\ Oppo,
Phys.\ Rev.\ A {\bf 66}, 053409 (2002)

\bibitem{PhaseoniumRev}
M.D. Lukin, P.R. Hemmer, and M.O. Scully, Adv. At. Mol. Opt. Phys.
{\bf 42}, 347 (2000).

\bibitem{Wilson-Gordon}
H. Shpaisman, A.D. Wilson-Gordon, and H. Friedmann, Phys. Rev. A
{\bf 70}, 063814 (2004); Phys. Rev. A {\bf 71}, 043812 (2005).

\bibitem{VandenHeuwell}
W.E. van der Veer, R.J.J. van Diest, A. Donszelmann, and H.B. van
Linden van den Heuvell, Phys. Rev. Lett. {\bf 70}, 3243
(1993).

\bibitem{Windholz96} W. Maichen, F. Renzoni, I. Mazets, E.
Korsunsky, and L. Windholz, Phys. Rev. A {\bf 53}, 3444 (1996).

\bibitem{Harris00}
A.J. Merriam, S.J. Sharpe, M. Shverdin, D. Manuszak, G.Y. Yin, and
S.E. Harris, Phys. Rev. Lett. {\bf 84}, 5308 (2000).

\bibitem{Windholz99}
E.A. Korsunsky, N. Leinfellner, A. Huss, S. Baluschev, and L.
Windholz, Phys. Rev. A {\bf 59}, 2302 (1999).

\bibitem{Windholz04}
A.F. Huss, R. Lammegger, C. Neureiter, E.A. Korsunsky, and L. Windholz,
Phys. Rev. Lett. {\bf 93}, 223601 (2004).

\bibitem{Sola05}
V.S. Malinovsky and I. R. Sola, Phys. Rev. Lett. {\bf 93}, 190502 (2004);
Phys. Rev. A {\bf 70}, 042304 (2004); Phys. Rev. A {\bf 70}, 042305 (2004).

\bibitem{AlkaliEarth}
R. Santra, E. Arimondo, T. Ido, C.H. Greene, and J. Ye,
Phys. Rev. Lett. {\bf 94}, 173002 (2005);
T. Ido, T.H. Loftus, M.M. Boyd, A.D. Ludlow, K.W. Holman, and J. Ye,
Phys. Rev. Lett. {\bf 94}, 153001 (2005).


\bibitem{QScully}
M.O. Scully and M.S. Zubairy, {\it Quantum Optics} (Cambridge
University Press, Cambridge, 1997).

\bibitem{AllenEberly} L. Allen and J.H. Eberly, {\it Optical Resonance
and Two-level atom} (Wiley, New York, 1975).

\bibitem{Footnote} For
$\alpha\to\infty$ this behavior becomes evident after rescaling the
field amplitudes, Eq.~(\ref{Gij}), with the linewidth $\gamma_e$
and then letting $\gamma_{jg}\to 0$.


\bibitem{Stroud76} R. M. Whitley and C. R. Stroud, Jr., Phys. Rev.
A {\bf 14}, 1498 (1976).

\bibitem{Boyd}
M.S. Malcuit, D.J. Gauthier, and R.W. Boyd, Phys. Rev. Lett. {\bf
55}, 1086 (1985); R.W. Boyd, M.S. Malcuit, D.J. Gauthier, and K.
Rza$\dot{\rm z}$ewski, Phys. Rev. A {\bf 35}, 1648 (1987).

\bibitem{Korsunsky02}
E.A. Korsunsky and M. Fleischhauer, Phys. Rev. A {\bf 66}, 033808 (2002).

\bibitem{Johnsson}
M.T. Johnsson and M. Fleischhauer, Phys. Rev. A {\bf 66}, 043808 (2002).

\end{thebibliography}
\end{document}